\documentclass[natbib209]{emulateapj}

\usepackage{amsmath}
\usepackage{epsfig}

\newcommand{\del}{\mbox{\boldmath $\nabla$}}
\newcommand{\curl}{\mbox{\boldmath $\nabla \times$}}

\newcommand{\cross}{\mbox{\boldmath $\times$}}

\newcommand{\vv}{{\bf v}}

\newcommand{\FF}{\mbox{\boldmath $F$}}
\newcommand{\BB}{\mbox{\boldmath $B$}}

\newcommand{\uvz}{\mbox{\boldmath $\hat{z}$}}
\newcommand{\uvt}{\mbox{\boldmath $\hat{\theta}$}}
\newcommand{\uvp}{\mbox{\boldmath $\hat{\phi}$}}

\shorttitle{Magneto-Shear Instabilities in the Tachocline}
\shortauthors{Miesch}

\begin{document}

\title{Sustained Magneto-Shear Instabilities in the Solar
Tachocline}

\author{Mark S.\ Miesch}
\affil{High Altitude Observatory, NCAR\altaffilmark{1}, Boulder, CO 80307-3000}
\email{miesch@ucar.edu}

\altaffiltext{1}{The National Center for Atmospheric Research is operated by 
the University Corporation for Atmospheric Research under sponsorship of 
the National Science Foundation}

\begin{abstract}
We present nonlinear three-dimensional simulations of the stably-stratified
portion of the solar tachocline in which the rotational shear is maintained
by mechanical forcing.  When a broad toroidal field profile is specified
as an initial condition, a clam-shell instability ensues which is similar
to the freely-evolving cases studied previously.  After the initial
nonlinear saturation, the residual mean fields are apparently too weak 
to sustain the instability indefinitely.  However, when a mean poloidal
field is imposed in addition to the rotational shear, a statistically-steady
state is achieved in which the clam-shell instability is operating continually.
This state is characterized by a quasi-periodic exchange of energy between
the mean toroidal field and the instability mode with a longitudinal wavenumber
$m=1$.  This quasi-periodic behavior has a timescale of several years and
may have implications for tachocline dynamics and field emergence patterns 
throughout the solar activity cycle.
\end{abstract}

\keywords{Sun: tachocline, MHD, Sun:rotation, Sun:interior}

\section{Introduction}\label{intro}

It is now well established (at least from a theoretical perspective)
that axisymmetric rings of toroidal field in the solar tachocline 
are susceptible to global instabilities induced by the 
latitudinal differential rotation.  This was first
demonstrated for 2--D spherical surfaces
\citep{gilma97,gilma99,dikpa99,gilma00,cally03,dikpa04} and was 
later extended to 3--D spherical shells under the shallow-water 
approximation \citep{gilma02,dikpa03} and the thin-shell
approximation \citep{cally03b,miesc07,gilma07}.  For strong fields 
(such that the magnetic energy is comparable to the kinetic energy
contained in the differential rotation), the most unstable modes
have longitudinal wavenumber $m=1$, which corresponds to a 
tipping of the ring such that its central axis becomes misaligned
with the rotation axis.

This class of global magneto-shear instabilities occurs both for 
concentrated bands of toroidal field and for broad field profiles.
The preferred mode for broad field profiles which are antisymmetric
about the equator is the clam-shell instability whereby toroidal 
rings in the northern and southern hemispheres tip out of phase, 
reconnecting at the equator on one side of the sphere and opening 
up on the other side \citep{cally01,cally03,miesc07}.  In the 
absence of external forcing, the instability proceeds until loops 
of field become perpendicular to the equatorial plane and the 
latitudinal shear is nearly eliminated.

However, the solar tachocline is not an isolated system.  Rotational
shear is continually maintained by stresses from the overlying convection
zone and the underlying radiative interior.  Meanwhile, magnetic flux
is continually being replenished from above by penetrative convection 
and meridional circulation.  If clam-shell instabilities are indeed 
occurring in the tachocline, their environment is likely much more
dynamic than the freely-evolving scenarios considered in previous 
nonlinear simulations \citep{cally01,cally03,miesc07}.

In this letter we present the first nonlinear simulations of 
global magneto-shear instabilities in the solar tachocline
in which the instabilities are continually maintained 
against nonlinear saturation and dissipation through the 
use of external forcing.  We use the same nonlinear, 3--D 
thin-shell model employed by \citet[][hereafter MGD07]{miesc07}
to study the freely-evolving case.  The linear stability of this 
thin-shell system has been investigated by \citet{gilma07}.
The reader is referred to these two papers for a much more comprehensive 
discussion of the unstable modes and their nonlinear saturation under
freely-evolving conditions.  These papers also contain a more comprehensive
discussion of related work and a more detailed description of the 
thin-shell model and the numerical algorithm.

The first question we address in this letter is whether the instabilities
can generate poloidal fields efficiently enough to act as a dynamo 
when only mechanical forcing of the rotational shear is applied.  
Our simulations suggest that this is unlikely although the results are 
inconclusive.  However, statistically steady states are found when the 
magnetic energy is maintained via a poloidal forcing term in the mean 
induction equation.  These states exhibit a quasi-periodic behavoir 
as mean fields continually transfer energy to the non-axisymmetric 
instability modes and are then re-established by the mechanical 
and magnetic forcing.

\section{The Thin-Shell Model}\label{model}
Our numerical model is based on the thin-shell approximation which
is described in detail by \citet{miesc04} and MGD07.  The aspect ratio
$\delta = D/R$ is assumed to be much less than unity, where $D$ and $R$
are the width and radial location of the computational domain.  Only terms
of lowest order in $\delta$ are retained in the equations of motion and
the layer is assumed to be stably-stratified, corresponding to the lower
portion of the solar tachocline \citep{thomp03}.  The equations are 
expressed in a rotating, spherical polar coordinate system defined by 
the colatitude $\theta$, longitude $\phi$ and height $z$ with 
corresponding unit vectors $\uvt$, $\uvp$, and $\uvz$.  The velocity 
and magnetic field components are defined as 
$\vv = u \uvp + v \uvt + \delta w \uvz$ and
$\BB = a \uvp + b \uvt + \delta c \uvz$.  The system
is made nondimensional through the use of horizontal and vertical 
length scales $R$ and $D$, a velocity scale based on the equatorial 
rotation rate, and the background density and entropy gradient.  We 
neglect pressure variations relative to entropy variations in the
linearized equation of state which corresponds to the Boussinesq
limit discussed by MGD07 [$\delta_P = 0$ in their equation (6)].
Apart from the parameters which specify the forcing, dissipation,
and initial conditions, this leaves one free parameter: the Froude
number $F_r$, which is a nondimensional measure of the stratification.
For the simulations discussed here we take $F_r = 0.1$ as appropriate
for the lower tachocline (MGD07).

The subgrid-scale (SGS) model consists of a fourth-order hyperdiffusion 
in the horizontal dimensions with an effective Reynolds number of $R_h$ 
and a Smagorinsky formulation for the vertical diffusion involving a 
vertical Reynolds number $R_v$ (MGD07).  The amplitude and form
of the SGS magnetic and thermal diffusion are assumed to be the
same as the viscous diffusion, implying magnetic and thermal Prandtl
numbers of unity.  The upper and lower boundaries of the layer 
are assumed to be impenetrable, stress-free, perfectly conducting,
and isentropic.

Mechanical forcing is imposed by adding a volumetric source term
to the momentum equation of the form
\begin{equation}\label{eq:Fu}
\FF_u(\theta,z) = \tau^{-1} \left(u_0 - \left<u\right>\right) \uvp
\end{equation}
where $u$ is the zonal velocity, $\tau$ is a characteristic timescale 
for the establishment of the shear and
\begin{equation}\label{eq:u0}
u_0(\theta,z) = \frac{1}{2} \left(1 - \cos\left(\pi z\right)\right)
\left(s_0 - s_2 \cos^2\theta\right) \sin\theta  ~~~. 
\end{equation}
Angular brackets $<>$ in equation (\ref{eq:Fu}) denote
an average over longitude.  In equation (\ref{eq:u0}), $s_0$ 
is the nondimensional equatorial rotation rate relative to the 
rotating reference frame and $s_2$ is the fractional angular 
velocity difference between the equator and poles.  Here we 
take $s_0 = 0.044$ and $s_2 = 0.18$ (MGD07).  The vertical 
profile of $u_0$ is such that the latitudinal shear is maximum 
at the top of the layer ($z = 1$) and vanishes at the bottom 
($z = 0$) in analogy with the solar tachocline.

In selected cases we also impose a dipolar field by means of a
similar forcing term added to the induction equation:
\begin{equation}\label{eq:FFp}
\FF_p(\theta,z) = \tau^{-1} \left(\BB_0 - \left<\BB_p\right>\right) 
\end{equation}
where $\BB_p = b \uvt + \delta c \uvz$ and
\begin{equation}\label{eq:B0}
\BB_0(\theta,z) = \curl \curl \left[C_0 \cos\theta \sin(\pi z) \uvz\right] ~~~.
\end{equation}
This field is intented to represent generated by dynamo processes in
the convection zone and pumped downward.
Imposing $\BB_0$ together with $u_0$ is an indirect way of imposing
a mean toroidal field; the latitudinal shear in the upper portion of
the layer stretches and amplifies the poloidal field through what is
known as the $\Omega$-effect, resulting in a mean toroidal field
$\left<a\right>$ which is antisymmetric about the equator.
The peak amplitude of $\left<a\right>$ will in general
depend on the poloidal field strength $C_0$, on the SGS 
diffusion, and to some extent on $F_r$.

\begin{figure}
\epsfig{file=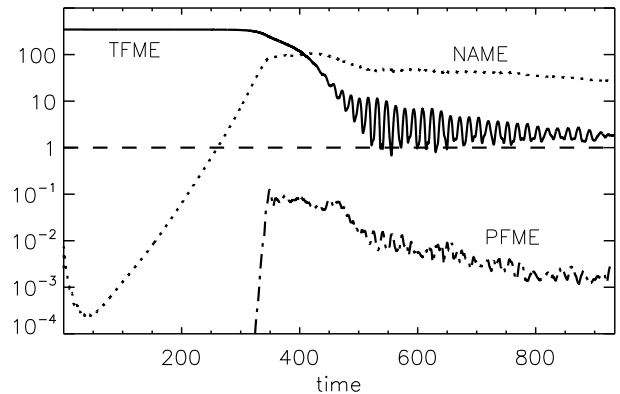,width=\linewidth}
\caption{The volume-integrated magnetic energy contained in 
the mean toroidal field (TFME, solid line), the mean poloidal 
field (PFME, dash-dotted line) and the non-axisymmetric field 
(NAME, dotted line) are shown as a function of time in a 
simulation with no magnetic forcing ($R_h = 10^7$, 
$R_v = 10^4$, $\tau=0.1$).  All curves are normalized relative 
to the integrated kinetic energy contained in the imposed 
shear profile $u_0$ (indicated by the horizontal
dashed line).\label{onlyshear}}
\end{figure}

\begin{figure}
\epsfig{file=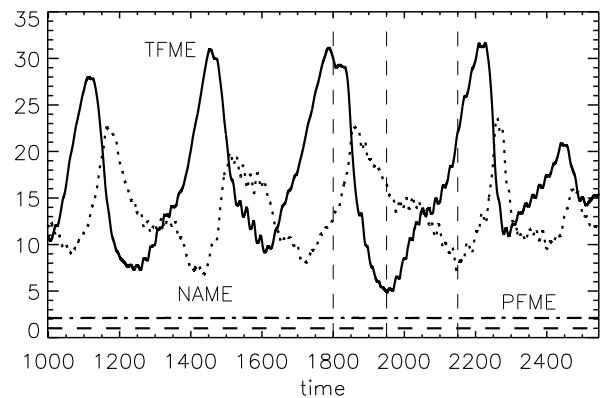,width=\linewidth}
\caption{As in Figure \ref{onlyshear} for a case
with both mechanical and magnetic forcing as
described in the text ($R_h = 10^5$, $R_v = 10^4$,
$\tau = 0.1$).  Note that the vertical axis is 
linear here rather than logarithmic.  Vertical 
dashed lines indicate the time instances 
illustrated in Figure \ref{slices}.\label{shearmag}}
\end{figure}

\begin{figure*}
\epsfig{file=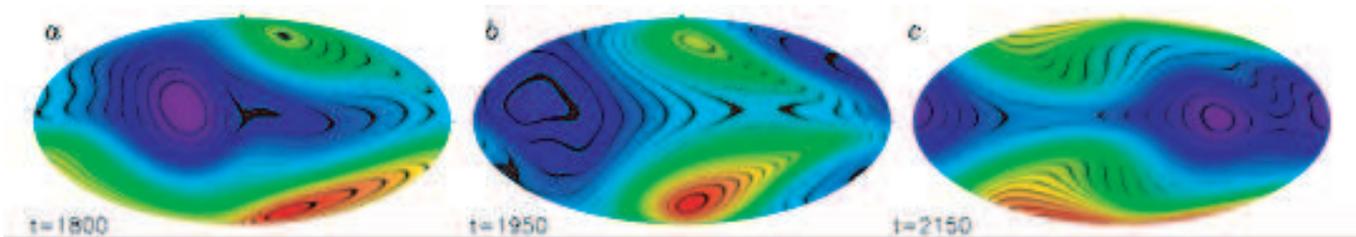,width=\linewidth}
\caption{The horizontal structure of the magnetic
potential $J$ near the top of the layer ($z = 0.98$)
is illustrated at three different times for the same
simulation as shown in Figure \ref{shearmag}.  The color
table varies with increasing $J$ as blue-green-yellow-red. 
Black countours trace horizontal magnetic field 
lines as noted in the text, although the sense of the
field is opposite in the northern and southern hemispheres.  
Each horizontal surface is displayed as a Molleweide 
projection in which lines of constant latitude appear 
horizontal.\label{slices}}
\end{figure*}

\section{Sustained Magneto-Shear Instabilities}\label{results}

Our simulations are initiated as described in MGD07, with
an equilibrium state defined by the zonal velocity $u_0$
and a broad toroidal field profile of the form
\begin{equation}\label{eq:Bphi}
a(\theta) = \alpha \cos\theta \sin\theta ~~~.
\end{equation}
The initial pressure and temperature are chosen such
that the initial state is in magneto-hydrostatic and
magneto-geostrophic balance.  The parameter $\alpha$ 
is set to unity for the simulations presented here, 
corresponding to a peak dimensional field strength of 
about 40 kG, toward the high end of the range expected 
to exist in the tachocline (MGD07).  Global magneto-shear
instabilities also occur for weaker fields but they take 
longer to develop.  The initial equilibrium state is 
perturbed by adding a random, small-scale velocity field.

Unlike the simulations presented in MGD07, we maintain
the differential rotation through the forcing term
defined in equation (\ref{eq:Fu}).  An example of the 
subsequent evolution in the absence of magnetic forcing 
($\FF_p = {\bf 0}$) is shown in Figure \ref{onlyshear}.  
The spatial resolution used for this case is $N_\theta$, 
$N_\phi$, $N_z$ = 128, 256, 210.

Figure \ref{onlyshear} shows the components of the magnetic
energy integrated over the volume of the shell.  One time
unit in the nondimensional system corresponds to about four days
(MGD07).  At early times the mean toroidal field (TFME) dominates 
but the non-axisymmetric component (NAME) grows rapidly as
the clam-shell instability develops.  Most of the
NAME is in the $m=1$ mode which dominates the total
magnetic energy after the instability saturates
at $t \sim 400$.

The evolution shown in Figure \ref{onlyshear} is very
similar to the unforced cases discussed at length
in MGD07.  However, in the absence of mechanical
forcing, the saturation of the clam-shell instability
induces a global redistribution of angular momentum
which reverses the sense of the differential rotation
(MGD07).  The forcing suppresses this, leaving the
differential rotation profile unchanged after 
saturation.

After saturation, the magnetic energy in the
mean toroidal field decreases as in the 
unforced case, despite the persistent 
rotational shear.  The TFME oscillates
with a period of about 15 time units, 
corresponding to about two months.  At later 
times, the amplitude of the oscillation decreases
and the period increases slightly, to about
18 time units.  This oscillation appears to 
be induced by a standing Alfv\'en wave excited
by the initial saturation of the instability.
However, low-wavenumber Rossby waves are also
excited and have a comparable period.

In shallow-water systems which have a deformable
upper boundary, global magneto-shear instabilities
can possess significant kinetic helicity, suggesting
they may serve to generate poloidal field from 
toroidal field and thus drive a self-sustained
dynamo contained entirely within the tachocline
\citep{dikpa01b,gilma02,dikpa03}.  

The simulation shown in Figure \ref{onlyshear} does not
extend more than a diffusive timescale 
($\sim 10^5$ time units) so it is uncertain 
whether or not it may be classified as a dynamo.
The energy in the mean fields may be levelling 
off beyond $t \sim 800$ but it is unclear whether
this state will persist indefinitely.  A more diffusive 
analogue of this case ($R_h = 10^5$) was certainly not 
a dynamo since the total magnetic energy decayed 
steadily after the initial saturation.

In any case, it appears either that the clam-shell 
instability is no longer operating beyond 
$t \sim 500$ or that it is operating on a much longer 
time scale.  The magnetic energy remains dominated 
by the $m=1$ component for at least five years in 
dimensional time units.  Over this time period the
energy in the mean toroidal field remains larger
than the equipartition energy associated with
the imposed differential rotation but the energy 
in the mean poloidal field is about three orders of 
magnitude smaller.  These fields appear to be 
a remnant of the initial saturation of the 
instability at $t \sim 400$ as opposed to 
dynamo-generated fields induced by ongoing 
instabilities.

The situation changes dramatically if a mean poloidal
field is imposed through the forcing term expressed
in equation (\ref{eq:FFp}).  Figure \ref{shearmag} 
illustrates the evolution of the magnetic energy 
components in a simulation with both mechanical
and magnetic forcing ($N_\theta = 64$, $N_\phi = 128$, 
$N_z = 210$). The time span shown covers a period beyond 
the initial saturation of the instability, after a 
statistically steady state has been reached.

The time evolution shown in Figure \ref{shearmag} reflects
a quasi-periodic exchange of energy between the mean toroidal
field and the non-axisymetric field components, the latter
dominated by the $m=1$ mode. The first two oscillations
shown each span about 340 time units which corresponds to
3.7 years.  However, the TFME drops lower in the subsequent
cycle, reaching a minimum at $t \sim 1950$, leading to a longer
cycle of about 400 time units (4.4 years).  The following 
cycle is then much shorter, lasting only about 230 time
units (2.5 years).  The shape of each cycle is asymmetric,
with a relatively slow rise in TFME followed by a sharper
drop as the clam-shell instability sets in.

The non-axisymmetric magnetic energy exhibits quasi-periodic
cycles similar to the mean toroidal field but phase-shifted
such that the maxima in NAME occur as the TFME is decreasing.
Again, this reflects the repeated development of the clam-shell
instability which transfers energy from the mean toroidal field 
to the $m=1$ components.  After the instability saturates,
the mechanical and magnetic forcing re-establish the mean fields
and the next cycle proceeds.

To gain further insight into the dynamics, it is 
instructive to express the magnetic field in terms of
scalar magnetic potentials $J$ and $C$ defined such that
\begin{equation}\label{eq:BJC}
\BB = \curl \left(J \uvz\right) + \curl \curl \left(C \uvz\right) ~~~.
\end{equation}
In the simulations reported here, as in the freely evolving cases
reported in MGD07, the magnetic field remains predominantly 
horizontal and the first term on the right-hand-side of 
equation (\ref{eq:BJC}) dominates. Thus, to a good approximation,
contours of $J$ trace the horizontal field lines 
as illustrated in Figure \ref{slices}
[$\curl (J \uvz) = \uvz \cross \del J$]. 

The changing patterns shown in Figure \ref{slices} illustrate
the competing effects of the forcing and the instabilities.
At $t = 1800$ (Fig.\ \ref{slices}$a$), the mean toroidal field
dominates the magnetic energy, although non-axisymmetric structure
is evident.  By $t = 1950$ the clam-shell instability has
transferred much of this energy to the $m=1$ mode and horizontal
field lines are oriented more north-south (Fig.\ \ref{slices}$b$).  
The imposed shear then operates on these fields as well as the 
imposed poloidal field to rebuild the mean toroidal field which 
again dominates by $t = 2150$ (Fig.\ \ref{slices}$c$).

The vertical structure of the flow is similar to analogous
freely-evolving cases with vertical shear discussed in MGD07.
The velocity and magnetic fluctuations remain predominantly
horizontal and the instability proceeds
most vigorously near the top of the layer where the latitudinal
shear is strongest.

For the simulation shown in Figures \ref{shearmag} and 
\ref{slices}, the amplitude of the imposed poloidal field, 
$C_0$, is such that the integrated magnetic energy PFME is 
about twice the equipartion value $u_0^2/2$ (integrated over 
the volume).  This 
ratio is somewhat smaller near the top of the layer where the 
latitudinal shear peaks but nevertheless it is probably 
unrealistically large for the solar tachocline.  Weaker
imposed fields could potentially produce mean toroidal 
fields of comparable strength but this can only be achieved in 
a simulation if the SGS diffusion is sufficiently low.
Indeed, an analogous simulation with a weaker 
imposed field (PFME/DRKE = 0.02) and the same diffusion coefficients 
($R_h = 10^5$, $R_v = 10^4$) produced weaker mean toroidal fields 
(TFME/DRKE $\sim$ 3).  This system also exhibits sustained clam-shell 
instabilities with quasi-periodic behavior but the phase relationship 
between the TFME and NAME is not as well defined as in Figure 
\ref{shearmag}.  Some quasi-periodic oscillations are present
on timescales of several years but longer-term trends are also
evident, comparable to or longer than the duration of the 
simulation ($\sim$ 10 years).  Such longer-term evolution is 
to be expected since the growth rate of the clam-shell 
instability decreases with decreasing field strength.

Achieving substantially lower SGS diffusion would require
higher resolution which is a challenge because of the 
long integration times necessary to capture multiple 
cycles.  The relatively short forcing timescale used
for the simulations shown in Figures \ref{onlyshear}-\ref{slices}
($\tau = 0.1$, corresponding to about 10 hours) is also
in some sense required by the setup of our numerical 
experiments.  In a simulation similar to that shown in
Figures \ref{shearmag}-\ref{slices} but with $\tau = 2$
(8 days), the mechanical forcing is insufficient to 
overcome the magnetic tension associated with the 
imposed poloidal field.  As a result, the DRKE is only
20\% of the target value associated with $u_0$ and the 
clam-shell instability is suppressed. The field does 
exhibit $m=1$ structure but TFME dominates the magnetic
energy and the evolution is quasi-steady, with a slow 
retrograde propagation and no episodic opening 
up of the clam-shell pattern.  

The solar tachocline is much more complex than our highly 
simplified model.  Forcing timescales are longer but the 
diffusion is lower, so strong toroidal fields could be 
produced with lower poloidal field strengths and the differential
rotation could be maintained with weaker mechanical
forcing.  It is therefore difficult to establish from
the simulations alone whether the solar tachocline is
in a regime which exhibits quasi-periodic behavoir as
in Figure \ref{shearmag}.  However, helioseismic inversions
indicate that the rotational shear in the tachocline is 
continually maintained and linear analysis suggests that
even weak toroidal fields are unstable in the presence
of such shear so it is likely that the tachocline is
indeed continually undergoing global magneto-shear 
instabilities.

Latitudinal shear in the convective envelope is maintained
by Reynolds stresses and meridional circulations on timescales
of order months while the nearly uniform rotation of the radiative
interior may be established on much longer time scales
\citep[e.g.][]{gough98,talon02,miesc05}.  Magnetic flux 
is continually supplied to the tachocline by topological 
pumping from penetrative convection \citep[e.g.][]{tobia01}.  Although 
much of this flux is disordered, mean fields may be generated by nonlinear 
self-organization processes such as inverse cascades or by 
rotational phase mixing and turbulent reconnection 
\citep{sprui99,brown06}.  In addition, global meridional 
circulations may transport axisymmetric poloidal flux to 
the tachocline where mean toroidal fields would then be 
generated by rotational shear \citep{dikpa04b,charb05}.

In this letter we have demonstrated that the clam-shell instability
can operate continually when rotational shear and magnetic fields
are continually replenished.  The temporal evolution is 
quasi-periodic and as mean fields alternately build up and destabilize.  
This has the character of a critical phenomenon but it is not
self-organized criticality in the technical sense because there
is a characteristic time and spatial scale associated with
the growth rate and wavenumber of the instability
\citep{jense98}.  It is likely that other instability modes
may similarly be maintained continually by external forcing.
Notable among these is the $m=1$ tipping instability which
exists for the type of concentrated toroidal bands thought 
to give rise to photospheric active regions
\citep{dikpa99,cally03,miesc07}.

If global magneto-shear instabilities are indeed occurring
in the solar tachocline they would have wide-ranging 
implications for tachocline dynamics and for the coupling
between the convective envelope and the radiative interior.
Angular momentum transport induced by these instabilities
may influence the differential rotation profile in the
convection zone and the longer-term rotational evolution 
of the Sun \citep{charb93,gilma00b,talon02}.  Chemical transport across 
the tachocline also has implications for solar evolution, helioseismic 
structural inversions, and photospheric abundance measurements 
\citep{christ02,pinso97}.  Tachocline
instabilities may also play a role in the solar dynamo.  Although
the simulations reported here suggest that clam-shell instabilities
may not be capable of sustaining a dynamo localized entirely
within the tachocline, they may still play a role in 
parity selection, field propagation, and flux emergence 
patterns \citep{gilma00b,dikpa01c,norto05,dikpa05}.

\acknowledgments
We thank Peter Gilman and Mausumi Dikpati for many informative 
discussions and for comments on the mauscript.  This work was 
supported by NASA under the work order W-10,177 and made use 
of high-performance computing resources at the National Center 
for Atmospheric Research.


\begin{thebibliography}{31}
\expandafter\ifx\csname natexlab\endcsname\relax\def\natexlab#1{#1}\fi

\bibitem[{Browning {et~al.}(2006)Browning, Miesch, Brun, \& Toomre}]{brown06}
Browning, M.~K., Miesch, M.~S., Brun, A.~S., \& Toomre, J. 2006, \apjl, 648,
  L157

\bibitem[{Cally(2001)}]{cally01}
Cally, P. 2001, Solar Physics, 199, 231

\bibitem[{Cally(2003)}]{cally03b}
Cally, P.~S. 2003, \mnras, 339, 957

\bibitem[{Cally {et~al.}(2003)Cally, Dikpati, \& Gilman}]{cally03}
Cally, P.~S., Dikpati, M., \& Gilman, P.~A. 2003, \apj, 582, 1190

\bibitem[{Charbonneau(2005)}]{charb05}
Charbonneau, P. 2005, Living Reviews in Solar Physics, 2, online journal,
  http://www.livingreviews.org/lrsp-2005-2 (cited Jan 2007)

\bibitem[{Charbonneau \& MacGregor(1993)}]{charb93}
Charbonneau, P. \& MacGregor, K.~B. 1993, \apj, 417, 762

\bibitem[{Christensen-Dalsgaard(2002)}]{christ02}
Christensen-Dalsgaard, J. 2002, Rev. Mod. Phys., 74, 1073

\bibitem[{Dikpati {et~al.}(2004{\natexlab{a}})Dikpati, Cally, \&
  Gilman}]{dikpa04}
Dikpati, M., Cally, P.~S., \& Gilman, P.~A. 2004{\natexlab{a}}, \apj, 610, 597

\bibitem[{Dikpati {et~al.}(2004{\natexlab{b}})Dikpati, de~Toma, Gilman, Arge,
  \& White}]{dikpa04b}
Dikpati, M., de~Toma, G., Gilman, P.~A., Arge, C.~N., \& White, O.~R.
  2004{\natexlab{b}}, \apj, 601, 1136

\bibitem[{Dikpati \& Gilman(1999)}]{dikpa99}
Dikpati, M. \& Gilman, P.~A. 1999, \apj, 512, 417

\bibitem[{Dikpati \& Gilman(2001{\natexlab{a}})}]{dikpa01b}
---. 2001{\natexlab{a}}, \apj, 551, 536

\bibitem[{Dikpati \& Gilman(2001{\natexlab{b}})}]{dikpa01c}
---. 2001{\natexlab{b}}, \apj, 559, 428

\bibitem[{Dikpati \& Gilman(2005)}]{dikpa05}
---. 2005, \apjl, 635, L193

\bibitem[{Dikpati {et~al.}(2003)Dikpati, Gilman, \& Rempel}]{dikpa03}
Dikpati, M., Gilman, P.~A., \& Rempel, M. 2003, \apj, 596, 680

\bibitem[{Gilman(2000)}]{gilma00b}
Gilman, P.~A. 2000, Solar Physics, 192, 27

\bibitem[{Gilman \& Dikpati(2000)}]{gilma00}
Gilman, P.~A. \& Dikpati, M. 2000, \apj, 528, 552

\bibitem[{Gilman \& Dikpati(2002)}]{gilma02}
---. 2002, \apj, 576, 1031

\bibitem[{Gilman {et~al.}(2007)Gilman, Dikpati, \& Miesch}]{gilma07}
Gilman, P.~A., Dikpati, M., \& Miesch, M.~S. 2007, \apjs, in press

\bibitem[{Gilman \& Fox(1997)}]{gilma97}
Gilman, P.~A. \& Fox, P.~A. 1997, \apj, 484, 439

\bibitem[{Gilman \& Fox(1999)}]{gilma99}
---. 1999, \apj, 510, 1081

\bibitem[{Gough \& McIntyre(1998)}]{gough98}
Gough, D.~O. \& McIntyre, M.~E. 1998, Nature, 394, 755

\bibitem[{Jensen(1998)}]{jense98}
Jensen, H.~J. 1998, Self-Organized Criticality (Cambridge: Cambridge Univ.\
  Press)

\bibitem[{Miesch(2005)}]{miesc05}
Miesch, M.~S. 2005, Living Reviews in Solar Physics, 2, online journal,
  http://www.livingreviews.org/lrsp-2005-1 (cited Jan 2007)

\bibitem[{Miesch \& Gilman(2004)}]{miesc04}
Miesch, M.~S. \& Gilman, P.~A. 2004, Solar Physics, 220, 287

\bibitem[{Miesch {et~al.}(2007)Miesch, Gilman, \& Dikpati}]{miesc07}
Miesch, M.~S., Gilman, P.~A., \& Dikpati, M. 2007, \apjs, in press

\bibitem[{Norton \& Gilman(2005)}]{norto05}
Norton, A.~A. \& Gilman, P.~A. 2005, \apj, 630, 1194

\bibitem[{Pinsonneault(1997)}]{pinso97}
Pinsonneault, M. 1997, \araa, 35, 557

\bibitem[{Spruit(1999)}]{sprui99}
Spruit, H.~C. 1999, \aap, 349, 189

\bibitem[{Talon {et~al.}(2002)Talon, Kumar, \& Zahn}]{talon02}
Talon, S., Kumar, P., \& Zahn, J.-P. 2002, \apjl, 574, L175

\bibitem[{Thompson {et~al.}(2003)Thompson, Christensen-Dalsgaard, Miesch, \&
  Toomre}]{thomp03}
Thompson, M.~J., Christensen-Dalsgaard, J., Miesch, M.~S., \& Toomre, J. 2003,
  \araa, 41, 599

\bibitem[{Tobias {et~al.}(2001)Tobias, Brummell, Clune, \& Toomre}]{tobia01}
Tobias, S.~M., Brummell, N.~H., Clune, T.~L., \& Toomre, J. 2001, \apj, 549,
  1183

\end{thebibliography}

\end{document}